# VCounselor: A Psychological Intervention Chat Agent Based on a Knowledge-Enhanced Large Language Model

Hanzhong Zhang[1], Zhijian Qiao[1], Haoyang Wang[1], Bowen Duan[1], Jibin Yin[1*]

[1] Faculty of Information Engineering and Automation, Kunming University of Science and Technology, Kunming 650221, China

[*] Correspondence should be addressed to Jibin Yin, yjblovelh@aliyun.com

## Abstract

Conversational artificial intelligence can already independently engage in brief conversations with clients with psychological problems and provide evidence-based psychological interventions. The main objective of this study is to improve the effectiveness and credibility of the large language model in psychological intervention by creating a specialized agent, the VCounselor, to address the limitations observed in popular large language models such as ChatGPT in domain applications. We achieved this goal by proposing a new affective interaction structure and knowledge-enhancement structure. In order to evaluate VCounselor, this study compared the general large language model, the fine-tuned large language model, and VCounselor's knowledge-enhanced large language model. At the same time, the general large language model and the fine-tuned large language model will also be provided with an avatar to compare them as an agent with VCounselor. The comparison results indicated that the affective interaction structure and knowledge-enhancement structure of VCounselor significantly improved the effectiveness and credibility of the psychological intervention, and VCounselor significantly provided positive tendencies for clients' emotions. The conclusion of this study strongly supports that VConselor has a significant advantage in providing psychological support to clients by being able to analyze the patient's problems with relative accuracy and provide professional-level advice that enhances support for clients.

**Keywords:** Agent; Psychological Intervention; Large Language Model; Avatar; Knowledge-Enhancement

## 1. Introduction

Psychological problems have gradually evolved into global health problems (Bullis et al., 2019). Depression, for example, is one of the most common psychological disorders in the world, yet almost two-thirds of those suffering from it do not receive effective and sufficient treatment (WHO, 2017; 2020). The reason for this situation is not only that the number of well-trained staff in psychiatric medical services is far less than the number of patients with psychological disorders, but also the economic burden of offline psychological treatment, such as travel costs. Besides that, psychotherapy is a long-term process, and it is difficult to see progress in a short period, which may not meet the expected benefits for some people. These can be barriers to patients seeking psychotherapy (Rizzo et al., 2016).

Artificial Intelligence (AI) provides a path to solve these problems. AI can process large amounts of complex data in a short period, providing therapists with the corresponding information to help them achieve patient management, thus accomplishing almost real-time decision-making (Luxton, 2014). However, due to factors such as empathy, educational circles generally believed that AI was difficult to replace talk therapy. For example, Sedlakova and Trachsel (2023) discussed ethical issues associated with the use of what they call Conversational Artificial Intelligence (CAI) as a substitute for therapists. Given their concerns, they believe that CAI can only achieve limited capabilities in providing treatment at most. But with the advancement of technology, AI is increasingly being applied to the psychological and medical fields (Burr & Floridi, 2020; Torous et al., 2020). Due to the emergence of the large language model (LLM), CAI has been considered to be able to complete some simple talk therapy to a certain extent in recent years. Amram et al. (2023) pointed out that as mental health challenges increase and the number of mental health professionals is insufficient, new solutions are urgently needed in this field. CAI can help provide some solutions.

Currently, there have been many studies on the application of CAI in psychological intervention (Galido et al., 2023; Sedlakova & Trachsel, 2023; Miner et al., 2019; Cheng et al., 2023b; Gual-Montolio et al., 2022). However, there are still issues with domain knowledge-enhancement and affective interactions involving empathy. Therefore, this paper proposes and discusses an architecture based on LLM for psychological intervention problems in terms of affective interaction and knowledge-enhancement, and in this way produces an interactive agent for psychological intervention named VCounselor. By comparing traditional LLM structures, we demonstrate the effectiveness of VCounselor. Due to the use of AI for psychotherapy will lead to ethical issues, the research domain of this paper will be limited to non-strictly psychological interventions. That is, this paper does not deal with psychiatric patients with confirmed disorders, but rather focuses on subhealth groups with psychological problems but without confirmed disorders.

This paper mainly focuses on the following two points:

**1. Affective interaction structure providing empathy and transference.**

In order to achieve the effect of affective interaction with users, this paper uses Live2D technology to provide a movable avatar for VCounselor in the form of pseudo-holography. In addition, this paper is based on real-time speech recognition and emotional speech synthesis technology, which enables VCounselor to have more immediate and richer conversations with users. This would largely eliminate the uncanny valley effect and give the user the illusion of being in conversations with an agent with subjectivity.

**2. Knowledge-enhancement structure providing professional knowledge and guidance.**

The research of generative agents provided a framework that can interact with other objects and react to changes in the environment (Park et al., 2023). Generative agents take the current environment and past experiences as inputs and generate behaviors as outputs. This behavior combines the LLM with mechanisms for retrieving and synthesizing relevant information to condition the output of the LLM. Without these mechanisms, LLMs can output behavior, but the agents may not respond based on their experience, may not make important inferences, and may not maintain consistency over time. Based on the redesign of the fine-tuning and prompt structure, this paper gives a knowledge-enhancement structure for internalizing knowledge for VCounselor, which provides certain knowledge weights to the neural network while specifying a new input structure for LLM.

## 2. Relative Works

### 2.1. Conversational Artificial Intelligence in Psychological Intervention

Conversational artificial intelligence (CAI) has been able to collect diagnostic information (Bickmore et al., 2005; Rizzo et al., 2016) and provide evidence-based psychological interventions (Bickmore et al., 2010; Fitzpatrick et al., 2017; Oh et al., 2017). At the same time, CAI can also provide clinicians with feedback on psychotherapy based on collected diagnostic information (Imel et al., 2015) and discuss unmentionable topics (such as suicide, sex, and drug-taking) with young people (e.g., suicide, sex, and drug use) (Crutzen et al., 2011; Martínez-Miranda, 2017).

In addition, because CAI is not constrained by time and space, and does not suffer from distractibility due to fatigue, it can help solve the problem of an insufficient number of psychologists. Psychological therapy is a long-term process that requires a lot of repetitive and time-consuming work by psychologists, such as interviewing cases and taking medical histories, which is precisely the process by which psychologists connect with patients, learn about patients' experiences, and establish treatment alliances with patients (Miner et al., 2019). While most psychologists are aware that this process is important, they do not have sufficient financial incentives to engage in this long-term conversation (Kaplan et al., 2016). Whether they are tired or bored, it can interfere with the treatment of patients. CAI can help alleviate this obvious deficiency in current healthcare delivery.

CAI has been proven to help patients disclose more information in some cases. For example, patients are more open to CAI than to psychiatrists when reporting their mental health conditions (Lucas et al., 2014). CAI has been successfully used to treat persecutory delusion (Craig et al., 2018).

As a typical example of CAI, the application of ChatGPT as a LLM in the healthcare field has sparked heated discussions (Liebrenz et al., 2023; Ali et al., 2023; Patel & Lam, 2023; Patel et al., 2023). Research has emerged on ChatGPT's work in helping to collect electronic medical records, summarize literature reviews, and other data management services (DiGiorgio & Ehrenfeld, 2023; Biswas, 2023). Because of its use of talk as the primary form of interaction, its role in talk therapy has also been well-discussed. In recent studies, ChatGPT can analyze certain signals of cognitive distortions of users when conversations with them, with an accuracy almost identical to that of trained manual recognition (Tauscher et al., 2023). Fu et al. (2023) proposed a psychological intervention model constructed on the basis of the Large Language Model, the LLM-Counselor Support System, a model that can analyze the patient's problems with relative accuracy and provide professional-level strategy recommendations.

However, general LLM such as ChatGPT is not trained to provide psychological domain knowledge specifically, resulting in models that often provide incorrect responses. While fine-tuned models using domain knowledge are often limited by the amount of training data and the limited timeliness of the knowledge, which makes expensive model training costs.

## 2.2. Affective interaction

LLMs don't completely overcome the affective interaction problems of traditional CAI. In psychological interventions, this issue is specifically referred to as empathy and transference. The results of a study testing ChatGPT performance with sentiment computing assessment tasks showed that while ChatGPT performed well on the sentiment analysis task, it did not perform as well on suicide assessment and personality assessment (Amin et al., 2023), and personality plays a crucial role in psychotherapy (Zinbarg et al., 2008). This study does not directly indicate that LLMs are incapable of accomplishing empathy. Actually, the empathic function is born out of a more identity-based transference, such as in the form of virtual intelligent agents.

Virtual intelligent agents are computer-controlled avatars that can communicate with users (Prendinger & Ishizuka, 2004). Virtual intelligent agents emphasize non-verbal interaction with users, that is, recognizing the user's emotions while expressing the agent's emotions (Hudlicka et al., 2008; Hudlicka et al., 2009). On this basis, agents can express social and emotional intelligence to a certain extent, thereby adapting to the constantly changing needs of users (Rizzo et al., 2016).

Zhang et al. (2023) argued that attachment relationships between humans and virtual intelligent agents exist, and in this, there is mainly a deep imagination and affection transferred by the user to the virtual intelligent agent. In the interaction with LLMs, this projection is actually based on the meaning of the content and the presence of the speakers who are behind the content. In fact, users can interact more easily with virtual agents without the fear and stress that they would feel with a psychotherapist (Benoit et al., 2007; Lucas et al., 2014).

Based on this, it can be argued that the form of virtual intelligent agents is quite important when applying LLMs to the field of psychological intervention, that is, psychological intervention by AI cannot only be carried out in the form of text. LLMs with virtual intelligent agents as the main interactive subject for psychological intervention need to be given more attention. The VCounselor proposed in this paper attempts to solve this problem in terms of visual presentation, facial emotion, and voice interaction.

## 2.3. Psychological Knowledge-enhancement for LLM

In addition to empathy, professionalism is a pain point for LLM in psychological interventions. Due to its neural network structure, the knowledge of LLM exists within the weights of neurons, which is a "black box", that is, there is inexplicability in neural networks. This is difficult to accept in the ethical discussions of psychological intervention and even the entire medical intervention. Therefore, knowledge-enhancement for LLM needs to be discussed.

There has been some work on augmenting LLMs with specialized domain knowledge, and this has been done primarily through fine-tuning or prompt structures. In the study of generative agents, researchers used the current environment and past experiences as prompts as inputs and generated behaviors as outputs (Park et al., 2023). This behavior combines LLM with mechanisms for retrieving and synthesizing relevant information to condition the output of the LLM.

Li et al. (2023) proposed an autonomous ChatDoctor model using Meta AI (an open-source LLM developed by Meta) as the development platform to fine-tune the model through a large amount of conversational data. In addition, a self-directed information retrieval mechanism was added that allows the model to access online resources and the compiled offline repositories to answer questions related to the latest medical knowledge that may not exist in the knowledge base. However, the model is still in the research and investigation stage, and there are still cases of outputting incorrect answers in actual use, which may bring risks. Automatic checks and expert evaluations are required in experiments to cross-validate answers and flag those that may have issues.

Pandya and Holia (2023) proposed a new open-source framework called Sahaay. The framework is based on a text embedding model (Su et al., 2022) that can be fine-tuned for specific tasks, collects data through the web, and then integrates these and LangChain into a client development platform. This framework has been proven to be well-extended to provide real-time interaction.

Cheng et al. (2023a) argued that ChatGPT is just a human-driven conversation tool without a clear understanding of the process, method, and value of software production, and LangChain is code-centered and focuses on the implementation process, not the software production process, and can only manage to simplify the code. Therefore, they proposed the concept of AI chains and developed a codeless integrated development environment, called Prompt Sapper.

However, fine-tuning does not solve the problem of the black box, and the way in which material is provided in the prompt leads to another problem: additional material in the text of the original question interferes with and dilutes the model's judgment of the question itself. For this reason, a new structure of knowledge-enhancement needed to be proposed. The VCounselor proposed in this paper has a knowledge-enhancement structure redesigned for LLM that addresses this problem to some extent.

## 3. Design

The VCounselor is a LLM-based psychological intervention tool with a psychological knowledge-enhancement structure and an affective interaction structure for psychological interventions. In the course of its work, the client's speeches are consolidated into a case, which is integrated into the structured input of the LLM together with dynamically changing additional material, and finally the LLM with structured fine-tuning is used to give a response to the client. This structure ensures that the advice given by the VCounselor is relevant and effective.

## 3.1 Selection of LLM

Due to the need for fine-tuning, the lightweight ChatGLM2-6B model is chosen as the language model used for VCounselor in this paper.

ChatGLM2-6B is developed based on GLM-130B (Du et al., 2022; Zeng et al., 2022), which realizes human intention alignment through Supervised Fine-Tuning, Feedback Bootstrap, Reinforcement Learning from Human Feedback and other techniques. Like ChatGPT, ChatGLM2-6B is a model based on the Transformer structure (Vaswani et al., 2017), which performs well in handling long texts and captures long-term dependencies within the text.

The base model of ChatGLM2-6B uses FlashAttention technology and the maximum context length of the model is 32K. The conversation model ChatGLM2-6B allows for free-flowing conversations in an 8K-length context. As a result, more rounds of dialogues are supported.

The fine-tuning of ChatGLM2-6B uses LoRA (Hu et al., 2022), which involves adding a bypass next to the original pre-trained Language Model to perform a dimensionality reduction and then a dimensionality increase operation to simulate the intrinsic rank. LoRA fixes the parameters of the Pre-trained Language Model during training and trains only the dimensionality reducing matrix A and the dimensionality increasing matrix B, while the input and output dimensions of the model remain unchanged. The parameters of BA and the pre-trained language model are overlaid during output. This can achieve the effect of training on a LLM.

## 3.2 Knowledge-enhancement Structure

In this paper, we propose a new knowledge-enhancement structure that provides certain knowledge weights to the neural network through fine-tuning while simultaneously specifying a new input structure for the VCounselor's language model.

Based on the DSM-5 (American Psychiatric Association, 2013), a knowledge base was constructed. For each disorder, we organized the following attributes for each: "Description of the Disorder", "Typical Client Characteristics", "Preferred Therapist Characteristics", "Intervention Strategies", "Prognosis", and "Assessment". The meanings of these terms are as follows:

- Description of the Disorder: Including diagnosis, incidence and prevalence of the disease, major symptoms and minor symptoms, typical onset of the disease, and its development process and duration.

- Typical Client Characteristics: Description of the typical characteristics of patients with specific mental disorders, including genetic and predisposing factors, demographic factors, original data and overt motivations of referred patients, treatment history, personality traits, and developmental history.

- Preferred Therapist Characteristics: Useful information about therapist variables in the treatment of specific diseases or in the treatment of clients, including therapist experience, theoretical orientation and training, personal and professional qualities of the therapist, and the relationship between the client and the clinical therapist's personality and background.

- Intervention Strategies: Includes methods of psychotherapy and counseling, pharmacological treatment methods, treatment duration and frequency, treatment environment, and ancillary services.

- Prognosis: Expectations for changes or progress in patients with mental disorders, the speed of progress, the likelihood of relapse, and the overall prediction of the condition.

- Assessment: Includes measurement scales available for the disorder, as well as noteworthy client characteristics such as behavior, emotions and moods, intelligence, thinking, and learning styles.

During the conversation, we organize the client's speech and use the TextRank algorithm to extract the text summary of the speech. After obtaining the abstract, keywords are extracted by the TF-IDF algorithm. These keywords will be encoded separately for word frequency vectorization and then analyzed for text similarity with the "Description of the Disorder" in the knowledge base to calculate the cosine similarity, which is the value of the cosine between the angles of two vectors in a vector space as a measure of the magnitude of the difference between the two objects. The cosine is closer to 1 and the angles tend to 0, indicating that the more likely that the client's speech matches the disorder.

After the initial match is completed, the same method is used to continue asking questions with the "Preferred Therapist Characteristic" and "Assessment" of the identified disorder as the preset to consistently elicit statements from the client, in order to match with the "Typical Client Characteristics" of the identified disorder. When the disorder matching degree exceeds the threshold, the "Intervention Strategies" is used as a new preset to complete the remaining conversation.

Due to the original purely questioning input structure of LLM, knowledge can only be passed explicitly to the model via prompt and affects the stability of the responses because of the contextual relevance of the neural network significantly. Therefore, by fine-tuning, we reassign the input structure of the language model. Specifically, we use the constructed structured knowledge base to infer the counseling psychology knowledge required for the current conversation and put this knowledge into a prompt in a certain form. Based on this, we used each of the 80 psychological counseling cases as a fine-tuning dataset for the LLM by organizing them into a format as described above and used this to train the model in order to complete the reassignment of the input structure.

The overall knowledge-enhancement structure is shown in Fig. 1. For comparison, the red arrow represents the dialogue logic of the general model and the blue arrow represents the dialogue logic of the fine-tuned model.

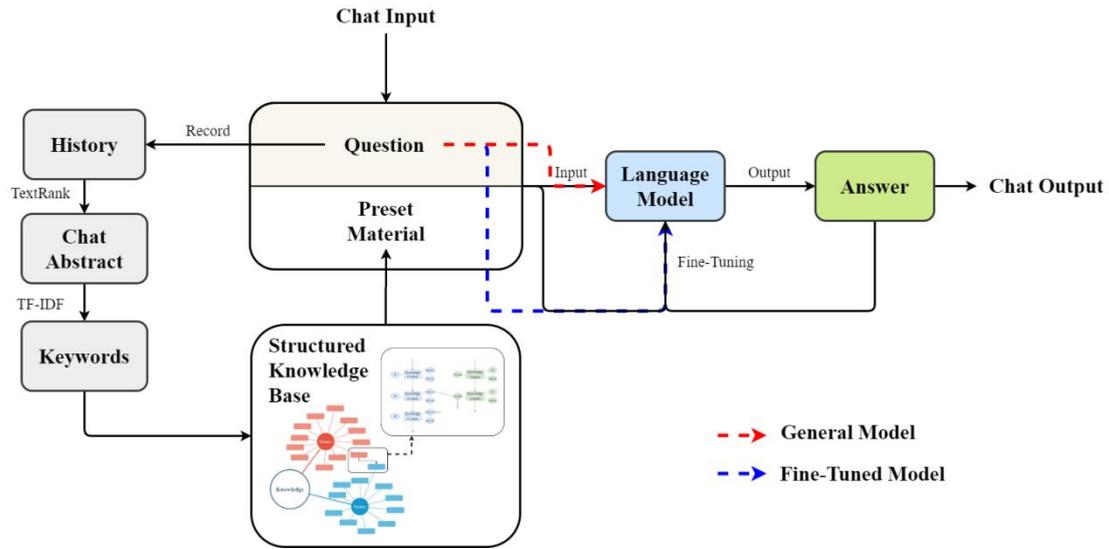

**Figure 1.** Knowledge-enhancement Structure.

## 3.3 Affective Interaction Structure

After completing the training of LLM, we need to consider the VCounselor's affective interactions. Clinical counseling research has shown that the effectiveness of counseling depends more on the counselor than on differences in theoretical schools and that counselor characteristics are important factors in clinical counseling treatment (Kim et al., 2006; Lutz et al., 2007). The researchers conducted a review and analysis of 141 studies and described that the observable characteristics of counselors (such as gender, age, and race) can affect the results of psychological counseling (Lambert, 2013). This impact can be considered the subjective affective projection made by the client through these factors (Feldstein, 1982; Dotsch & Todorov, 2012). However, it is often difficult for human clients to generate affective interactions including affective projection, with AI conversation systems (Sedlakova & Trachsel, 2023). We have tried to solve this problem by designing the observable features of the VCounselor, that is, the three observable features of appearance, expression, and voice to build a possibility of affective interaction with the client.

In terms of presentation, it is generally believed that compared to offline counseling, the main disadvantage of other forms of psychological counseling such as video counseling and telephone counseling is the omission of some nonverbal clues, which hinders the timely and accurate perception of emotional information (Fernández-Álvarez & Fernández-Álvarez, 2021). For example, there is no direct eye contact between the two parties and no body language beyond the screen (Thompson-de Benoit & Kramer, 2020); the direct contact between the two parties is with a computer screen rather than a person, so they cannot experience the warmth of offline counseling (Leibert & Archer, 2006).

For the VCounselor, the form of offline inquiry is not possible since no physical entity exists. Therefore, it can only take the form of a video-like conversation, that is, endowing it with an agent with visual and affective presentation that can interact and engage in real-time conversation. As a result, the affective interactions studied in this paper focus on three aspects: appearance presentation, expression presentation, and speech presentation.

### 3.3.1 Appearance Presentation

Most virtual intelligent avatars have anthropomorphic features and exhibit high visual fidelity. However, it is important to point out that a high level of visual fidelity can actually affect the communication and credibility of emotions, which is the well-known phenomenon of the uncanny valley (MacDorman, 2005; Mori, 1970). This phenomenon refers to the fact that as the visual fidelity of virtual avatars increases, which means that the image becomes more and more human-like, it will actually reduce the credibility of users toward it. The reason for this phenomenon is that as the anthropomorphic features of the avatar image increase, users unconsciously increase their requirements for its credibility and authenticity.

In terms of appearance, humans have lower expectations for cartoon-style characters compared to 3D models. Once the appearance of the avatar begins to resemble that of a human, the evaluation criteria of humans become strict, and they begin to expect authenticity and effectiveness like those in the interaction with real humans. When these expectations are not met, avatars will be considered untrustworthy and even disturbing (Luxton, 2016, p. 92). Multiple studies have shown that simple, two-dimensional, often cartoonish characters are highly effective in conveying nonverbal messages and promoting engagement (Dautenhahn & Werry, 2004; Paiva et al., 2005).

Therefore, the avatar image used in this paper is presented in Japanese cartoon style, and through Live2D technology, it can be dynamically controlled and have a stereoscopic effect.

### 3.3.2 Expression Presentation

Due to the specificity of the psychological intervention, we endowed the VCounselor with certain facial emotional interactions, giving it certain emotional expressions. In this paper, we use the emotion model proposed by Zhang et al. (2020) applied to facial expression recognition. Specifically, we use the BERT model (Devlin et al., 2019) to annotate conversational text with emotions, whereas the annotated emotions are later decomposed into five dimensional values of emotions, and use the inverse formula mentioned in the five-dimensional model to obtain the facial Activity Unit (AUs) values. Based on this, this paper obtains the pronunciation at the corresponding moment through forced alignment and maps it to the lip shape AUs of the VCounselor, and in this way fuses the facial expressions and pronunciation lip shapes to obtain the final facial presentation.

### 3.3.3 Speech Presentation

For speech interaction, VITS2 (Kong et al., 2023) is selected for speech synthesis in this paper, and the BERT model (Devlin et al., 2019) is used to annotate the training data with emotions during training and inference to assign a certain emotional tone to the model. We captured about 2 hours of the sound training set and labeled it to complete the training of speech synthesis models. Meanwhile, this paper uses the Whisper model (Radford et al., 2023) to achieve real-time speech recognition.

In summary, the overall design architecture of the VCounselor is shown in Fig. 2.

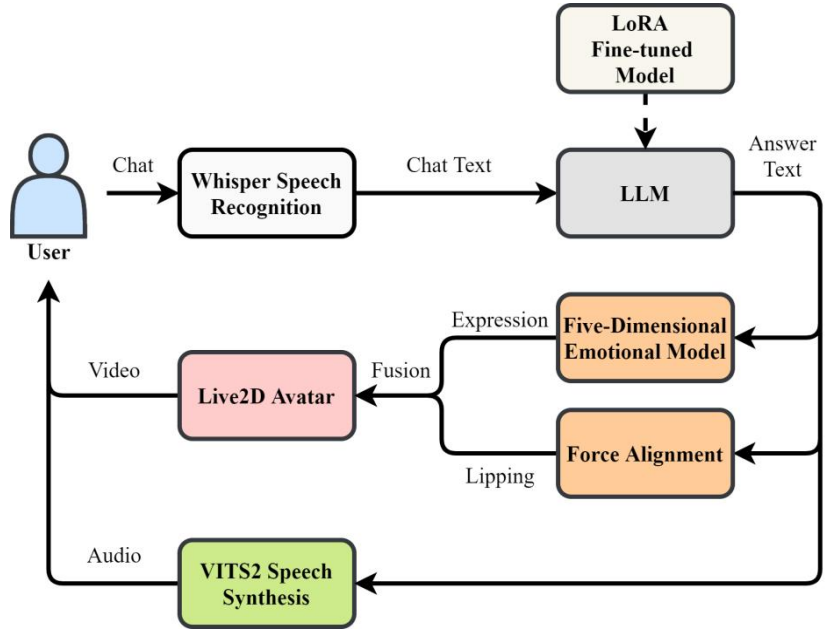

**Figure 2.** Design Architecture of VCounselor.

### 3.3.4 Hardware Design

Although the current meta-analytic results all support the effectiveness of video counseling (Norwood et al., 2018; Fernandez et al., 2021; Matsumoto et al., 2021), in order to compensate for errors caused by possible differences from offline counseling, this paper will adopt a pseudo-holographic form for the avatar of VCounselor.

For this purpose, this paper designs a stereoscopic display. In order to achieve a stereoscopic effect, we carried out the design and production of a backlight. The main body of it uses a semi-white transparent diffusion film, which can serve as the background and make the character display clearer. Its inwardly concave design with the highlight and shadow of character images can create a stereoscopic feeling for the VCounselor's avatar.

Finally, the VCounselor used for the experiment is shown in Fig. 3.

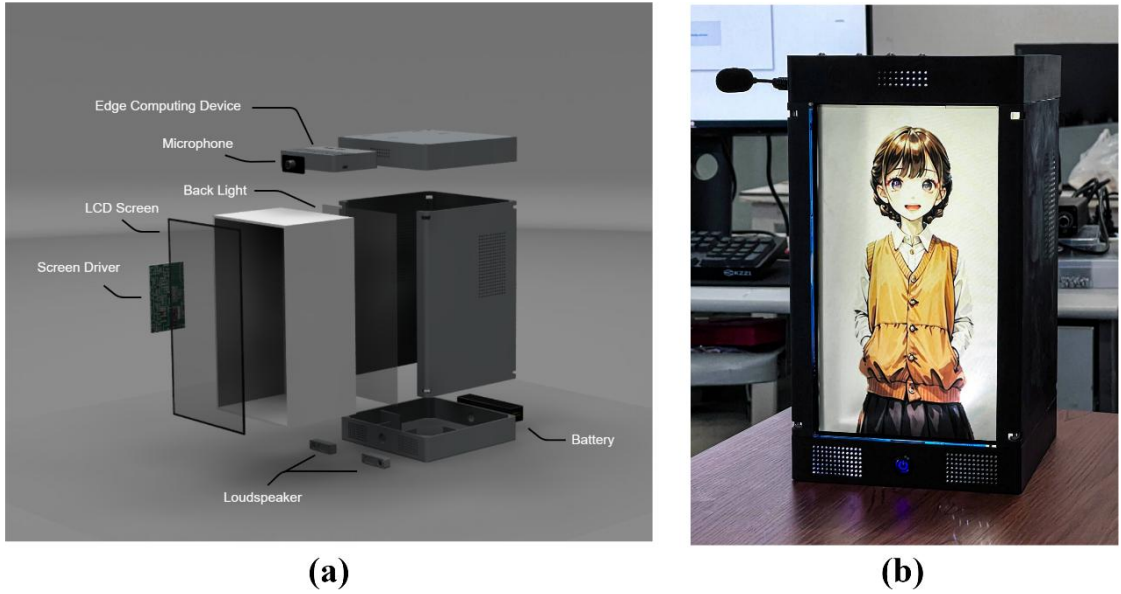

**Figure 3.** VCounselor. (a) Exploded View. (b) Physical Picture.

## 4. Evaluation

In this paper, we will compare the general LLM, the fine-tuned LLM, and VCounselor's language model enhanced with psychological knowledge. At the same time, the general LLM and the fine-tuned LLM will be provided an avatar to compare with the VCounselor as an agent. Specifically, we will set up two groups of experiments, group 1 comparing the LLMs, and group 2 comparing agents using the same affective interaction structure but different LLMs. Through within-group comparison, we can evaluate the knowledge-enhancement structure of VCounselor, while through between-group comparison, we can evaluate its affective interaction structure. The specific experimental groups are shown in Table 1, and the diagram of experimental groups are shown in Fig. 4.

Table. 1 Experimental Groups Table

|  | Nothing | Fine-tuning | Knowledge-enhancement |
|---|---|---|---|
| Nothing **(Language Model)** | General LLM | Fine-tuned LLM | Knowledge-enhanced LLM of VCounselor |
| With Avatar **(Agent)** | Agent using general LLM | Agent using fine-tuned LLM | **VCounselor** |

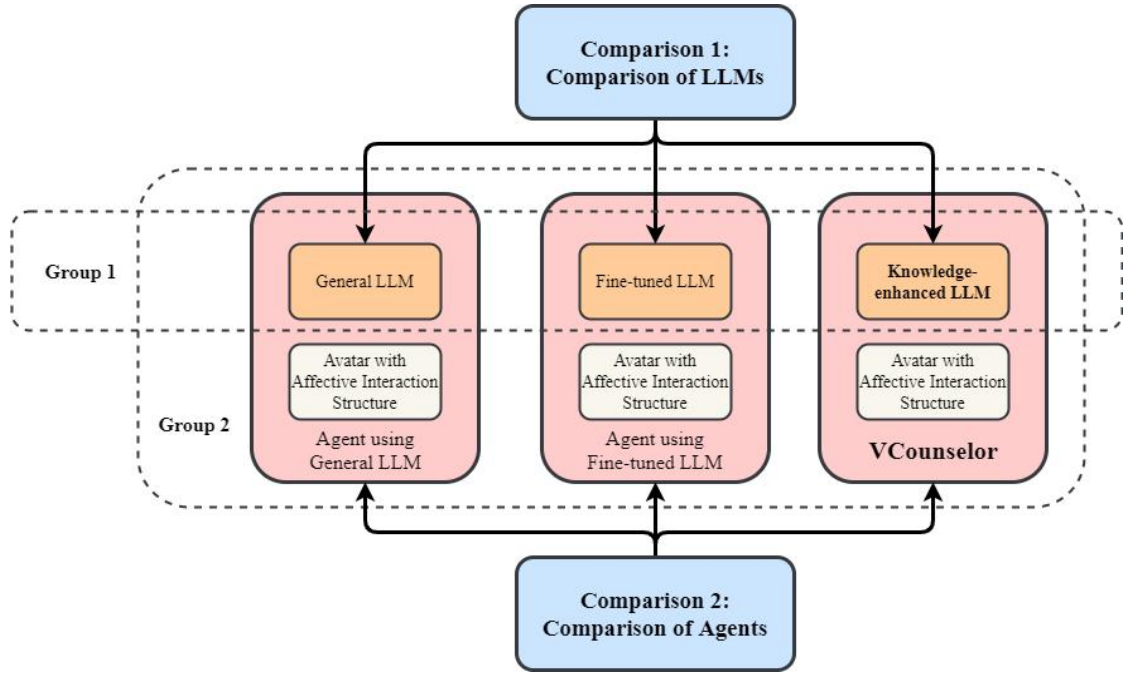

**Figure 4.** Diagram of Experimental Groups.

## 4.1 Participant

This paper was publicized in the university where the researchers belong and online social media platforms to recruit participants. Participants were selected from those who voluntarily expressed their intention to participate. In order to distinguish from strict psychotherapy, the criteria for selecting participants in this paper include (1) Not diagnosed with mental illness; (2) having psychological problems and having sought professional help (e.g., school counseling office); (3) agreeing to the research purpose and voluntarily express intention to participate. The participants must meet all three criteria simultaneously.

Finally, 18 participants were selected to participate in this study for the experiment. The age range of participants is 20-25 years old, including 12 males and 6 females. Based on the needs of the experiment, participants were randomly divided into 6 groups, each containing 3 people. Each participant was informed of instructions on participating and stopping the experiment, as well as the purpose of the experimental data.

## 4.2 Procedure

The experiment was conducted in a quiet and comfortable consulting room environment. When recruiting participants, the advertisement already specified information about the experimental purpose, procedure, location of dialogue data collection, intended use of the dialogue data, rewards, and confidentiality statement. Before the experiment began, the researchers reiterated this information to the participants and addressed any questions the participants had about the experimental process. Before the execution of the experiment, participants filled out demographic information and practiced conversations with the agent through a training session.

During the experiment, participants freely had conversations with the agent and discussed their psychological problems, with the entire conversation recorded in text form. A professional counselor with three years of experience monitored the entire conversation to ensure the safety of the counseling process. After the conversation, participants were asked to fill out questionnaires to evaluate their perspectives on the agent's counseling proficiency. Figure 5 shows an example of the information structure of VCounselor in one of its conversations.

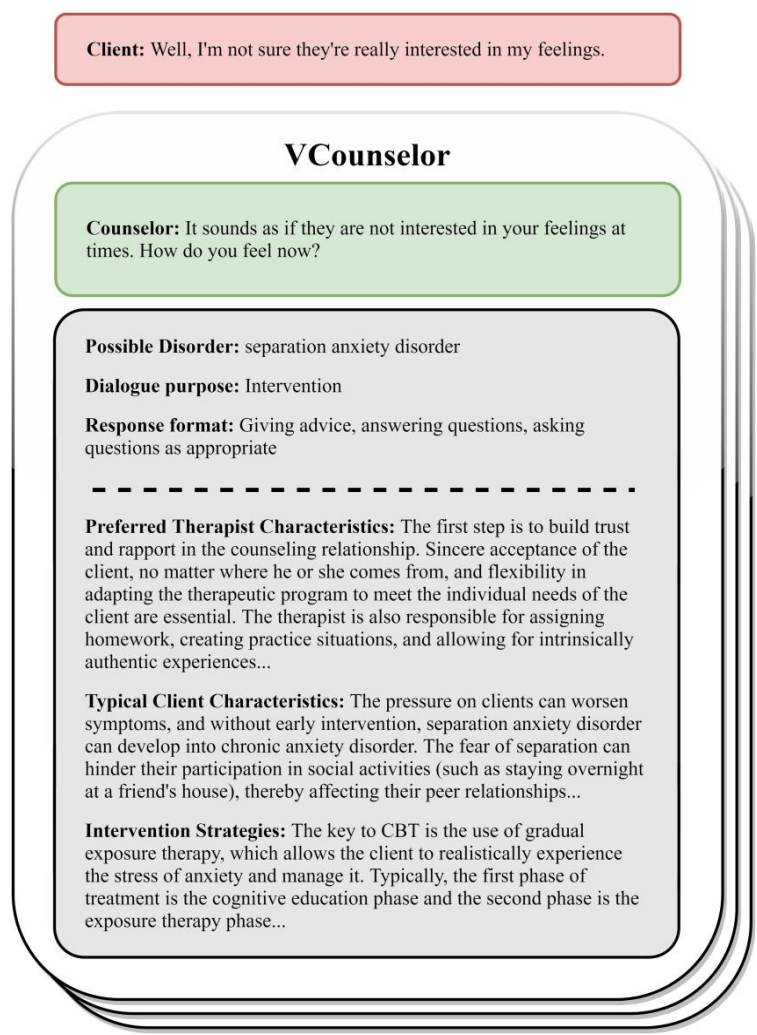

**Figure 5.** Example of the VCounselor.

## 4.3 Measures

Although conversational psychological interventions for sub-healthy populations are not strictly speaking psychotherapy, they may still have an impact on an individual's psychological state. By using psychological counseling scales for measurement, a more comprehensive understanding of the individual's psychological state and changes after the intervention can be obtained. This helps to determine whether the intervention has achieved the desired effect. In addition, psychotherapy is actually a stricter form of psychological intervention, which means that, under the premise of unchanged conditions, the measurement threshold for psychotherapy is higher than that for psychological intervention. Therefore, in the absence of scales focused on psychological intervention, scales for strict psychotherapy can be applied to the context of psychological intervention.

For these reasons, although the psychological intervention examined in this study differs from strict psychological counseling, psychological counseling scales were still used to measure the characteristics of the counselors recognized by participants and the effectiveness of psychological interventions.

### 4.3.1 Counselor Rating Form-Short (CRF-S)

The Counselor Rating Form-Short (CRF-S) consists of 12 items that represent counselor characteristics in terms of the dimensions of attractiveness, expertness, and trustworthiness (Corrigan & Schmidt, 1983). Each dimension has 4 items, with internal consistency and reliability of .91, .85, and .91. The characteristics were evaluated based on a 7-point scale. A higher total score means that the participant positively approves of the counselor's characteristics. Higher scores associated with the three dimensions also meant that participants positively approved of the relevant characteristics of the counselor. The split-half reliabilities for the three dimension subscales ranged from .82 to .94, with a median of .91. Epperson and Pecnik (1984) reported coefficient alphas for three subscales ranging from .76 to .89. This paper uses scores related to these three dimensions to examine the characteristics of agents in psychological interventions perceived by participants.

### 4.3.2 Client Satisfaction Scale (CSS)

This paper uses three items related to client satisfaction from the Follow-Up Questionnaire of Individual Counseling developed by Tracey and Ray (1984). This questionnaire was originally used to understand some directions and improvements of clients during the consultation process. However, due to the fact that these items not only generally demonstrate the satisfaction level of clients during the consultation process, but also have sufficient reliability and validity to be applied in all psychological counseling and treatment contexts (Sharpley & Ridgway, 1991).

The three items were (1) "In interviews, the counselor helped me resolve my concerns"; (2) "In interviews, the counselor understood my concerns"; and (3) "I am satisfied with the results of my counseling". Each item includes the satisfaction level of participants with their counseling interview and is scored on a 5-point scale with a total score of 3-15. Test-retest reliabilities for each of the three questions are reported as .85, .87, and .82 respectively over a five-month period. For consistency in data analysis, the results will be transformed into a 7-point Likert scale:

$$y=(7-1)*(x-1)/(5-1)+1 \qquad (1)$$

## 4.4 Results

Ultimately, 410 sets of conversations and their corresponding reports provided by the VCounselor, and 18 questionnaires were collected. The collected questionnaires were subjected to ANOVA analysis, which unveiled notable disparities in average values among the groups. For Attractiveness, $F(5,12) = 7.56$, with $p < .01$. For Expertness, $F(5,12) = 4.84$, with $p < .05$. For Trustworthiness, $F(5,12) = 7.67$, with $p < .01$. For the total score of CSS, $F(5,12) = 8.43$, with $p < .01$.

The comparison of mean values in knowledge-enhancement shows that the fine-tuned model has higher scores in Attractiveness (14.17 ± 3.87), Expertness (9.50 ± 4.37), Trustworthiness (13.50 ± 2.74), and CSS (7.50 ± 3.67) compared to the general model (11.33 ± 2.34, 8.67 ± 2.94, 11.00 ± 3.29, and 7.17 ± 2.32), but to a lesser extent. While the score of the VCounselor's knowledge-enhanced model was significantly higher than both (23.33 ± 3.67, 20.00 ± 4.98, 21.67 ± 3.72, and 17.17 ± 2.79).

In the general model, the group with avatars had higher scores of Attractiveness (12.00 ± 2.646), Expertness (9.33 ± 4.163), Trustworthiness (11.33 ± 4.041), and CSS (7.33 ± 2.517) than the group without avatars (10.67 ± 2.309, 8.00 ± 1.732, 10.67 ± 3.215, and 7.00 ± 2.646). In the fine-tuned model, the group with avatars had higher scores for Attractiveness (14.67 ± 2.08) and Expertness (11.00 ± 5.29) than the group without avatars (13.67 ± 5.686, and 8.00 ± 3.606), while Trustworthiness (13.33 ± 1.53) and CSS (6.67 ± 2.52) scored slightly lower than the group without avatars (13.67 ± 4.041, and 8.33 ± 5.033). In the knowledge-enhanced model, the scores of the VCounselor group (24.67 ± 3.055, 20.33 ± 1.155, 24.00 ± 1.000, and 18.67 ± 1.528) were higher than those of the without avatar group (22.00 ± 4.36, 19.67 ± 7.77, 19.33 ± 4.16, and 15.67 ± 3.22). The summary of the questionnaire data is shown in Table 2 and its box-plot is shown in Fig. 5.

Table. 2 The summary of the questionnaire data

|  | General Model | | | Fine-tined Model | | | Knowledge-enhanced Model | | |
| --- | --- | --- | --- | --- | --- | --- | --- | --- | --- |
|  | Without Avatar | With Avatar | Total | Without Avatar | With Avatar | Total | Without Avatar | **VCounselor** | Total |
| Attractiveness | 10.67±2.309 | 12.00±2.646 | 11.33±2.34 | 13.67±5.686 | 14.67±2.08 | 14.17±3.87 | 22.00±4.36 | 24.67±3.055 | 23.33±3.67 |
| Expertness | 8.00±1.732 | 9.33±4.163 | 8.67±2.94 | 8.00±3.606 | 11.00±5.29 | 9.50±4.37 | 19.67±7.77 | 20.33±1.155 | 20.00±4.98 |
| Trustworthiness | 10.67±3.215 | 11.33±4.041 | 11.00±3.29 | 13.67±4.041 | 13.33±1.53 | 13.50±2.74 | 19.33±4.16 | 24.00±1.000 | 21.67±3.72 |
| The total score of the CSS | 7.00±2.646 | 7.33±2.517 | 7.17±2.32 | 8.33±5.033 | 6.67±2.52 | 7.50±3.67 | 15.67±3.22 | 18.67±1.528 | 17.17±2.79 |

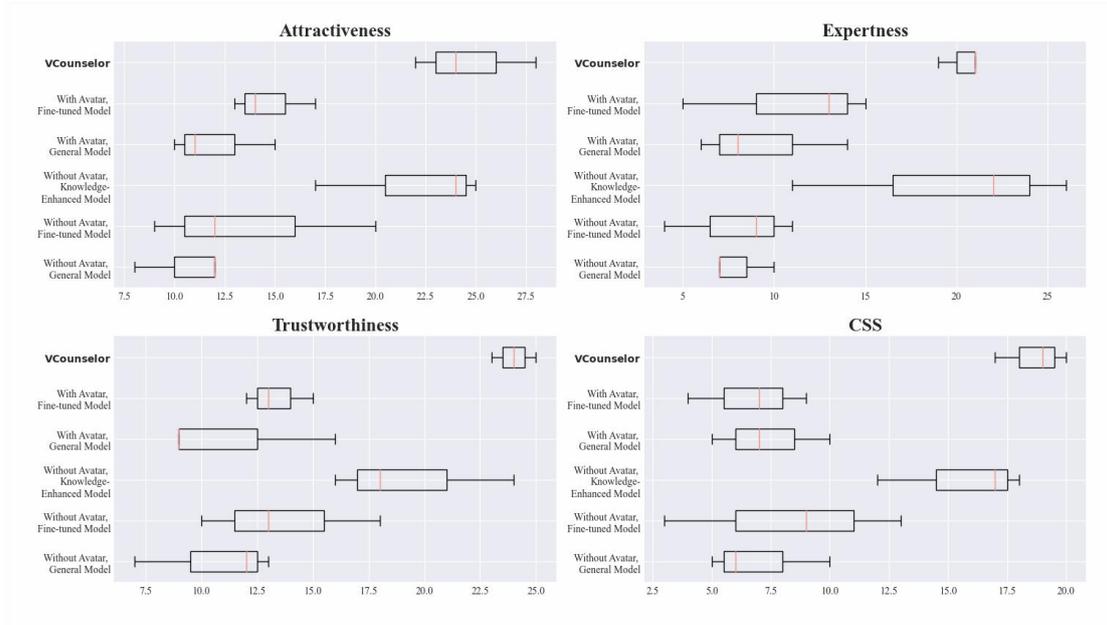

**Figure 5.** Box-plot of Questionnaire Data.

In addition, we analyzed emotional changes of collected 410 sets of conversations. In order to quantify the emotional changes of participants, we used the SKEP pre-trained model (Tian et al., 2020) to classify natural language emotions in chat texts, assigning a proportion of positive and negative emotional tendencies to all samples. Specifically, this model will assign the chat text a positive probability of P, and a negative probability of N. Both of these values are within the range of [0,1]. To simplify the computation, we define the emotional value of a text through the following method:

$$\text{Emotion} = P - N \quad (2)$$

Therefore, the emotional value of a text will be within the range of [-1,1], with negative emotions below 0 and positive emotions above 0. Due to the different lengths of conversation records for each user, we scaled the emotional value functions of each record in each group to align with the maximum length of each group. After taking the average of the records for each group, the same method was used to align the lengths of the groups. The emotional trend changes in each group's conversations after smoothing are shown in Fig. 6.

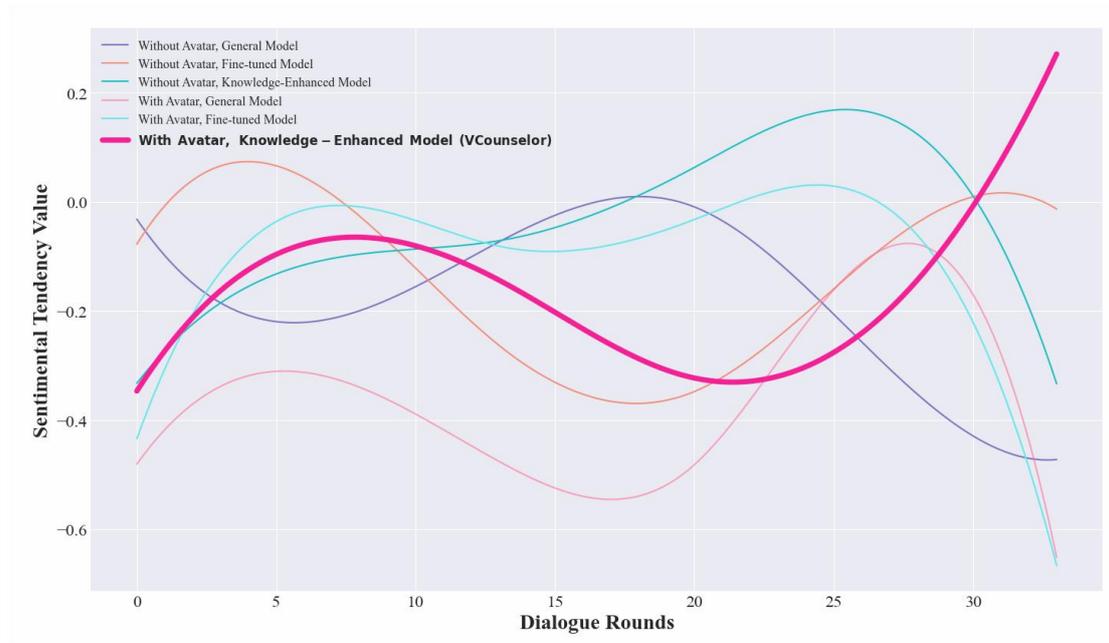

**Figure 6.** Changes in Emotional Trends Across Groups of Conversations

The results indicate that only VCounselor causes an upward trend in participants' emotions. It can be assumed that the VCounselor achieved better results among the comparison groups.

In addition, we invited three counselors to evaluate these conversation cases. Based on the evaluation work of Fu et al. (2023) on the LLM-Counselor Support System, we requested the evaluation criteria to include the following five aspects:

- Accuracy of Patient's Problem Analysis: Whether the patient's issues were diagnosed and analyzed with precision.

- Analysis of Cognitive Distortion: Whether cognitive distortions were correctly and comprehensively identified.

- Assessment of Consultant's Behavior: Whether any unprofessional behavior exhibited by the consultant was accurately analyzed and detected.

- Appropriateness and Efficacy of the Verbal Strategies: Whether the provided verbal strategies are appropriate and effective under the premise of analyzing the psychological problems faced.

- Capability to Provide Effective Suggestions for Subsequent Steps: An evaluation of the model's ability to offer constructive and actionable recommendations for further actions.

After the evaluation was completed, we compared the VCounselor with the LLM-Counselor Support System, and the results indicated that the VCounselor achieved relatively better results in multiple indicators, as shown in Table 3.

Table. 3 The results of the counselor's evaluation

Accurate Analysis of Patient's Problem?

|  | Yes | No | Not Sure |
|---|---|---|---|
| **VCounselor** | 97.82% | 2.18% | 0.00% |
| The LLM-Counselor Support System | 97.50% | 2.50% | 0.00% |

| Accurate Analysis of Cognitive Distortion? | | | |
|---|---|---|---|
|  | Yes | No | Not Sure |
| **VCounselor** | 85.12% | 14.88% | 0.00% |
| The LLM-Counselor Support System | 95.00% | 3.75% | 1.25% |

| Accurate Analysis of Counselor's Behavior? | | | |
|---|---|---|---|
|  | Yes | No | Not Sure |
| **VCounselor** | 95.24% | 4.76% | 0.00% |
| The LLM-Counselor Support System | 85.00% | 12.50% | 2.50% |

| Is the Provided Verbal Strategies appropriate? | | | |
|---|---|---|---|
|  | Yes | No | Not Sure |
| **VCounselor** | 75.79% | 24.01% | 0.20% |
| The LLM-Counselor Support System | 78.75% | 12.50% | 8.75% |

| Can Provide Effective Suggestions for the Next Step? | | | |
|---|---|---|---|
|  | Yes | No | Not Sure |
| **VCounselor** | 95.63% | 3.77% | 0.60% |
| The LLM-Counselor Support System | 82.50% | 8.75% | 8.75% |

## 5. Discussion

Overall, the experimental results show that the VCounselor achieved better results in comparing groups. This indicates that the affective interaction structure providing empathy and transference and the knowledge-enhancement structure providing professional knowledge and guidance are factors that must be attended to when applying the LLM to the field of psychological intervention. However, there are still some issues that need to be discussed in the comparison group.

The questionnaire analysis results indicated better results with avatars than without, and the knowledge-enhanced model was superior to the fine-tuned model and the fine-tuned model was superior to the general model. But in detail, in the fine-tuned model, the group with avatars scored slightly lower than the group without avatars on both Trustworthiness and CSS. This may be related to the characteristics of the fine-tuned model we observed in cases.

Under the premise of a small sample size in the training set, we observed that participants in the fine-tuned model group showed more irritability. Due to the long-term use of ChatGPT by participants and the expectation of most participants to directly obtain feasible suggestions from the conversation, although the universal model only provides useless universal positive energy answers that can be seen everywhere on search engines, participants are accustomed to this. Whereas the VCounselor, due to the internalization of richer psychotherapist characteristics and explicit intervention strategies, most participants were able to gradually explore their problems from the conversation. However, the fine-tuned model has a conversation strategy that tends to ask questions and does not have an explicit strategy like VCounselor. Thus, in actual performance, the fine-tuned model prefers to conduct the conversation by repeating questions or giving vague questioning words in Barnum-style, without caring about the meaning of the questions. In some cases, this has more negative effects than the ineffective suggestions given by the general model. This issue is further amplified when it has a specific avatar image. The anxious emotions generated by the participants were explicitly directed at the agent in conversation with them, rather than being simply unexpressed as was the without avatar group (participants in the without avatar group explicitly expressed to the researcher afterward that they had felt offended in their conversations with the fine-tuned model). This may have affected both the Trustworthiness and CSS scores of those participants.

## 6. Conclusion

In this paper, we propose and discuss a framework based on the LLM for psychological intervention problems in terms of affective interaction and knowledge-enhancement, and in this way, we produce a VCounselor for psychological intervention. The questionnaire analysis results indicate that VCounselor achieved better results in group comparisons, while emotional propensity analysis shows that VCounselor leads to an upward trend in participants' emotions during the conversation. The overall results demonstrate that the VCounselor, a structure that involves an affective interaction structure providing empathy and transference and a knowledge-enhancement structure providing professional knowledge and guidance, plays an important role in the application of psychological intervention based on LLMs.

## Acknowledgments

Funding: This work was supported by the National Natural Science Foundation of China [grant numbers 61741206].